\documentclass[aps,pra,twocolumn,showpacs,floatfix]{revtex4}

\usepackage{graphicx}% Include figure files
\usepackage{dcolumn}% Align table columns on decimal point

\usepackage{bm}
\usepackage{epsfig}
\usepackage{psfig}

\usepackage{epsf}
\def\inseps#1#2{\def\epsfsize##1##2{#2##1} \centerline{\epsfbox{#1}}}

\begin{document}

\title{Vortex lattices
in the lowest Landau level for confined Bose-Einstein condensates}
\author{ N.R. Cooper, S. Komineas}
\affiliation{Theory of Condensed Matter Group,  Cavendish Laboratory,
Madingley Road, Cambridge CB3 0HE, United Kingdom}
\author{N. Read}
\affiliation{Department of Physics, Yale
University, P.O. Box 208120, New Haven, CT 06520-8120.}

\date{\today}

\begin{abstract}
We present the results of numerical calculations of the groundstates of
weakly-interacting Bose-Einstein condensates containing large numbers of
vortices.  Our calculations show that these groundstates appear to be close to
uniform triangular vortex lattices. However, slight deviations from a uniform
triangular lattice have dramatic consequences on the overall particle
distribution. In particular, we demonstrate that the overall particle
distribution averaged on a lengthscale large compared to the vortex lattice
constant is well approximated by a Thomas-Fermi profile.

\end{abstract}

%03.75.Kk Dynamic properties of condensates; collective and hydrodynamic
%excitations, superfluid flow
%05.30.Jp Boson systems (for static and dynamic properties of
%Bose-Einstein condensates, see 03.75.Hh and 03.75.Kk)

\maketitle

\section{Introduction}

Experiments on rotating Bose condensates at high angular momentum have reached
the limit in which the vortex cores overlap strongly \cite{schweikhard}. In
this limit, the single particle states are restricted to states in the lowest
Landau level (LLL) \cite{wgs,butts}. The vortices cannot be considered to
interact by pairwise interactions, but have intrinsically multi-vortex
interactions \cite{TesanovicX91}, leading to a very interesting regime of
vortex physics.  

In an influential theoretical paper studying the nature of the vortex
lattices in this regime \cite{ho}, the assumption was made that the
vortices will form a uniform triangular lattice. Under this
assumption, it was shown that the particle density averaged over a
lengthscale large compared to the intervortex spacing has a Gaussian
profile.  One can, however, expect that in the true LLL groundstate the
vortices will not form an ideal triangular lattice, and that small
changes in vortex position could lead to a rather different overall
density profile; within simple approximate considerations, when there
are a large number of vortices one would expect these deviations to
lead to a Thomas-Fermi profile for the average particle
distribution \cite{footnote,shm2,watanabe}, which for a harmonic trap
is an inverted parabola as a function of the distance from the
rotation axis. 

We expect that the LLL approximation is an excellent guide to the
low-energy properties (including the ground state density profile)
whenever the ratio, $\lambda$, of the interaction energy scale to the
level spacing of the transverse harmonic confinement in an axially
symmetric trap, $\lambda \equiv 4\pi\hbar^2a_s\bar{n}/(M
\hbar\omega_\perp)$, is much smaller than unity ($\bar{n}$ is the
typical number density of bosons, $a_s$ the $s$-wave scattering
length, $M$ the particle mass, and $\omega_\perp$ is the trapping
frequency perpendicular to the rotation axis). We term $\lambda\ll 1$
the LLL regime. Note that the condition for being in this regime does
not involve the system size. When many vortices are present, it is
equivalent to the healing length being larger than the vortex
spacing. There will be perturbative corrections to the LLL
approximation, in powers of the interaction parameter $\lambda$. An
earlier work \cite{baym,watanabe}, which extends into the LLL regime,
finds within a variational ansatz that even if one neglects the
deformations of the vortex lattice that we study here, a TF profile
emerges through weak Landau level mixing provided the healing length
is small compared to the sample size.  Also Ref.~\cite{sheehy}
provides a treatment of the averaged density profile in the vortex
lattice in the opposite limit in which the healing length is small
compared to the vortex spacing, and deviations of the vortex lattice
from uniform triangular were found there.

In this paper, we provide explicit numerical calculations of the
groundstates of large numbers of vortices in atomic Bose condensates
in the LLL limit. By comparing our full variational results for the
many-vortex groundstate with the results obtained under the assumption
of a uniform triangular vortex lattice, we show that the small changes
in the vortex position in the true groundstate away from a uniform
lattice have a dramatic effect on the overall density profile. In
particular, we demonstrate that the coarse-grained particle density of
a system containing a large number of vortices is well approximated by
a Thomas-Fermi profile. The accuracy of the Thomas-Fermi profile in
describing this regime can already be seen from the results of the
numerical studies of Sinova {\it et al.}~\cite{shm2} where the
coarse-grained particle density for a small vortex lattice array in
the LLL was found to fit an inverted parabola.

\section{Model}
 
We consider the rotating Bose gas when interactions are sufficiently
weak to enter the LLL regime ($\lambda\ll 1$) and, furthermore, to
allow restriction to the two-dimensional (2D) limit in which all
particles occupy the lowest harmonic oscillator state of the axial
confinement (accurate provided $4\pi\hbar^2a_s\bar{n}/(M
\hbar\omega_\parallel)\ll 1$, where $\omega_\parallel$ is the trapping
frequency parallel to the rotation axis).  Although these conditions
are far from being satisfied in atomic gases in the absence of
rotation, at high angular momentum the radial expansion of the cloud
due to the centrifugal forces reduces the mean particle density and
leads the system towards the weakly-interacting
limit \cite{ho,baym}. Indeed, recent experiments \cite{schweikhard} have
reached conditions that are within the LLL regime and close to the
2D limit.

In this paper we shall restrict attention to the case where the
filling fraction (the ratio of the number of particles to the number
of vortices \cite{cwg}) is sufficiently large that the groundstate is a
vortex lattice \cite{cwg,shm1} and can be well-described by
Gross-Pitaevskii theory.  In the case of interest -- weak contact
interactions -- Gross-Pitaevksii theory for the groundstate amounts to
minimizing the interaction energy
\begin{equation}  \label{eq:psi4}
 E_{\rm int}=  \frac{2\pi\hbar^2 a_s N^2}{M}\int |\psi(\bm{r})|^4 \,d^3 \bm{r}
\end{equation}
within the space of states in the lowest subband of the axial confinement
($z$) and in the lowest Landau level of the transverse ($x-y$) motion.  
Such states may be written
\begin{equation}  \label{eq:psiv}
 \psi(\bm{r}) = A \prod_{j} (\eta-\eta_j) e^{-|\eta|^2/(2a_\perp^2)}
 e^{-z^2/(2a_\parallel^2)}
\end{equation}
where $\eta = (x + i y)$, and $a_{\parallel,\perp} \equiv
\sqrt{\frac{\hbar}{M\omega_{\parallel,\perp}}}$ are the oscillator
lengths parallel and perpendicular to the rotation axis. We choose the
normalization constant $A$ such that
\begin{equation}  \label{eq:norm}
\int
|\psi(\bm{r})|^2 d^3 \bm{r} = 1\, ,
\end{equation}
which is why the number of particles $N$ enters in the expression for the
interaction energy, Eqn.(\ref{eq:psi4}).
The variables $\eta_j$ are the complex numbers representing the
positions of the vortices in the $x-y$ plane.  Hence, the wave
function (including therefore the particle density) is fully specified
by the positions of the vortices.  The task is to find the positions
of the vortices that minimise the interaction integral as a function
of the total angular momentum per particle in units of $\hbar$,
\begin{equation}
\label{eq:angular}
\ell =  \int d^3{\bm r} \;\left[-i\psi^* \, {\bm r}\times\frac{\partial \psi}{\partial
{\bm r}}\cdot \hat{{\bm z}}
\right] \,. 
\end{equation}
Note that, within the LLL states, the combined kinetic and potential
energy (relative to the zero point energy) is proportional to the
angular momentum, which is why these energies need not be included
explicitly in the minimization when angular momentum is constrained.

The connection between vortex positions and particle density lies at
the heart of the difficulty of an exact derivation of the density
profile.  Variations in the coarse-grained 2D density profile,
$\bar{n}_{\rm 2d}(x,y)$, (the particle density $|\psi({\bm r})|^2$
averaged over lengthscales large compared to the vortex lattice
constant and integrated over the $z$ direction) are by necessity tied
to variations in the vortex positions.  (A similar relation of
superfluid density [which is the same as particle density in the
Gross-Pitaevskii theory] and vortex density as in the LLL
approximation \cite{ho} was found in Ref.~\cite{sheehy}.) If one
makes the assumption that the vortices are only slightly perturbed
from a triangular lattice, and ignores the variation of the lattice
geometry in the energetics, then one is led to the expectation that
the coarse-grained density profile will be an inverted parabola --
{\it i.e.} Thomas-Fermi like \cite{watanabe}. By imposing the
normalisation condition (\ref{eq:norm}), and noting that for states in
the lowest Landau level, the angular momentum per particle is directly
related to the density profile via
\begin{equation}
\label{eq:ell}
\ell = \int dx dy \left[\frac{(x^2+y^2)}{a_\perp^2} -1\right] \bar{n}_{2d}(x,y) \,
\end{equation} 
we find that the Thomas-Fermi profile may be written
\begin{equation}
\label{eq:tf}
\bar{n}_{\rm 2d}(\rho) =  \frac{2}{3\pi (\ell+1)a_\perp^2}\left[1 - \frac{\rho^2}{3(\ell+1)a_\perp^2}\right]
\end{equation}
where $\rho \equiv \sqrt{x^2+y^2}$. Thus, the radius of the cloud,
$R$, increases with angular momentum per particle as
\begin{equation}
R = \sqrt{3(\ell +1)}\; a_\perp \,.
\end{equation}
One can also relate the radius to the angular rotation frequency
$\Omega$ at which the state with angular momentum $\ell$ is
stable. Following the approach of Ref.~\cite{watanabe} one finds
\begin{equation}
R = \left[\frac{2 b g_{\rm 2D} a_\perp^2}{\pi \hbar (\omega_\perp - \Omega)}\right]^{1/4} \, ,
\end{equation}
where $b = 1.1596$ is the Abrikosov parameter for the energy of a
uniform triangular vortex lattice \cite{abrikosov} and $g_{\rm 2D} =
2\sqrt{2\pi} N\hbar^2 a_s/(a_\parallel M)$.

On the other hand, if one assumes that the vortices form a uniform triangular
lattice, then the coarse-grained 2D density profile is of the form \cite{ho}:
\begin{eqnarray}
\label{eq:gaussian}
\bar{n}_{\rm 2d}(\rho) & = & \frac{1}{\pi \sigma^2} e^{-\rho^2/\sigma^2}\\
\frac{1}{\sigma^2} & = & \frac{1}{a_\perp^2} - \pi n_V
\end{eqnarray}
where $n_V$ is the areal density of vortices. (For a triangular lattice of
lattice constant $a$, the vortex density is $n_V = 2/(\sqrt{3}a^2)$.)
Using (\ref{eq:ell}), one finds 
\begin{equation}
 \frac{1}{\sigma^2} = \frac{1}{a_\perp^2}\frac{1}{\ell+1}\,.
\end{equation}

For the full variational study, a numerical conjugate gradient method
is used to minimise the interaction energy (\ref{eq:psi4}) with
respect to the vortex positions in (\ref{eq:psiv}) and subject to the
normalisation (\ref{eq:norm}) for a range of values of the fixed
angular momentum per particle (\ref{eq:angular}).  To this end, it is
helpful to expand the wavefunction (\ref{eq:psiv}) in terms of the
single-particle states in the LLL with angular momentum quantum number
$m$, and write the (normalised) wavefunction in the $x-y$ plane as
\begin{equation}
\label{eq:cm}
\phi(\eta)  = \sum_{m\geq 0} c_m \frac{\eta^m}{\sqrt{\pi m!}\, a_\perp^{m+1}} e^{-|\eta|^2/(2a_\perp^2)} \,.
\end{equation}
Constraints on angular momentum per particle and normalisation are
then imposed by Lagrange multipliers (in terms of physical parameters
these are $\hbar(\omega_\perp -\Omega)N$ and $\mu N$, where $\mu$ is
the chemical potential relative to the zero-point energy).  In this
way, the variational equations can be expressed in terms of coupled
equations for the complex coefficients $c_m$.

As an aside, we note that the resulting variational equations can be
expressed in real-space form as
\begin{eqnarray}
\nonumber
g_{\rm 2D} \int d^2\eta' {{\delta}(\eta,\eta')} |\phi(\eta')|^2\phi(\eta')
- \mu \phi(\eta)  & & \\
  + \hbar(\omega_\perp-\Omega)\int \!\! d^2\eta' {{\delta}(\eta,\eta')} \frac{\eta\bar{\eta}' }{a_\perp^2}
\phi(\eta') 
& = &  0
\end{eqnarray}
where \begin{equation} {{\delta}(\eta,\eta')} = \frac{1}{\pi
a_\perp^2}\, e^{\left(-|\eta|^2/2a_\perp^2 -|\eta'|^2/2a_\perp^2
+\eta\bar{\eta}'/a_\perp^2\right)}\end{equation} is (the integral
kernel of) the projection operator to the LLL \cite{project}.  Notice
that $|\delta(\eta,\eta')| ^2 = e^{ - |\eta-\eta'|^2/a_\perp^2} / (\pi
a_\perp^2)^2$ falls rapidly for $|\eta-\eta'| > a_\perp$.  This
real-space form of the lowest Landau level variational equations is a
useful starting point for further analytic approximations.

Here we focus on the numerical solution to the problem. In our
numerical procedure we work in terms of the coefficients $c_m$, which
we allow to be non-zero up to some maximum angular momentum ($\leq
300$) which we check is sufficiently large to have no significant
effect on the results (this is equivalent to setting a maximum number
of vortices in (\ref{eq:psiv})).  This procedure is identical to that
followed in Ref.~\cite{butts}; the difference is that here we
are interested in much larger values of the angular momentum per
particle than the cases studied in Ref.~\cite{butts}.  (For the
Thomas-Fermi profile, at large angular momentum, one expects the
number of visible vortices, $N_V$, to increase with the angular
momentum per particle as $N_V = 3\ell$.)

\section{Results}

The results of our full variational calculation always show a
(slightly distorted) vortex lattice. We focus attention on the case of
large angular momentum where the lattice spacing appears to be roughly
uniform in the central region of the condensate.  For the range of
angular momentum that we study, we find that there are stable
low-energy configurations of the system in which one vortex is close
to the centre of the trap, and around which there is an approximate
6-fold rotation symmetry. One can then choose one of the lines of
reflection symmetry to be the $x$ axis.  We have tested that such
configurations are stable to small deviations around the minimising
values, and that they give the lowest energy that we could obtain.

\begin{figure}
\epsfig{file=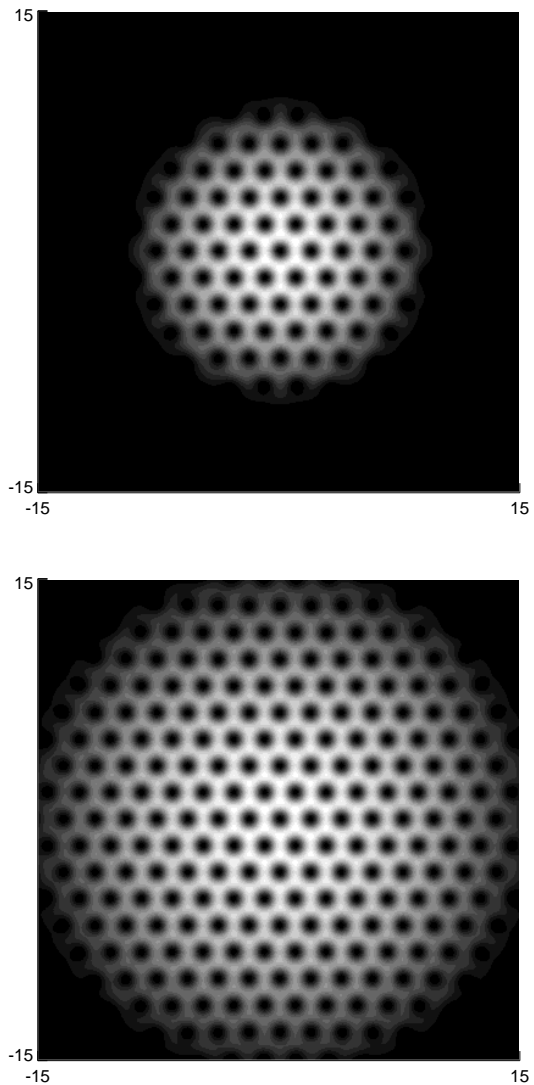,width=6cm}
  \caption{\label{fig:1} The particle density for two vortex lattices which
are calculated as the minimum of the interaction energy (\ref{eq:psi4}).  In the upper
entry we have angular momentum $\ell=31$ and in the lower entry $\ell=91$.
The grey-scale code is set to black for vanishing particle density and to
white for maximum density which occurs at the central region of the trap.  }
\end{figure}

In Fig.~\ref{fig:1} we show shadow plots of the particle density for
two examples of the results of the full variational study, at
$\ell=31$ and $\ell=91$.  Even though the vortex positions of the full
variational groundstate appear to be very close to the positions of a
triangular lattice, there are slight deviations from this regular
arrangement, especially close to the edge of the cloud. These
deviations are sufficient to give a very different overall density
profile to that expected from a uniform vortex lattice.
\begin{figure}
\epsfig{file=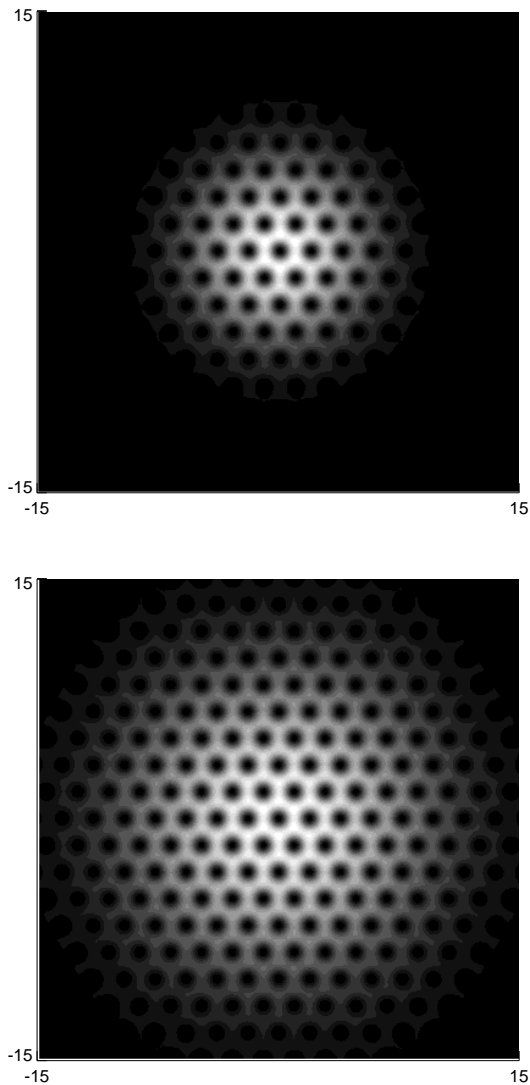,width=6cm}
 \caption{\label{fig:2} The particle density for two wavefunctions obtained
   under the assumpion of a perfect triangular vortex lattice.  In the upper
   entry we use a lattice constant $a=1.9351 a_\perp$ which corresponds to angular
   momentum $\ell=31$ and in the lower entry we have $a=1.9151 a_\perp$ which gives
   $\ell=91$.  The grey-scale code is set to black for vanishing particle
   density and to white for maximum density which occurs at the central region
   of the trap.  A qualitative comparison with Fig.~\ref{fig:1} is easy but
   note that the white colour in Fig.~\ref{fig:1} does not correspond to the
   same value in the present figure.  }
\end{figure}

To illustrate this, we have constructed the wavefunctions for the ansatz in
which the vortices are at the sites of a uniform triangular lattice \cite{ho}.
To most closely reproduce our full variational results we choose to position
the lattice such that there is a lattice site at the centre of the condensate
and there is reflection symmetry in the $x$-axis.  The only remaining freedom
in the ansatz is then the value of the lattice constant, which controls the
angular momentum per particle.

In Fig.~\ref{fig:2} we show the particle densities for this triangular
lattice ansatz, with lattice constant chosen to give angular momenta
$\ell=31$ and $\ell=91$.  The differences between the particle
densities obtained from the full variational calculation and from the
triangular lattice ansatz are large enough to be seen directly from
the shadow plots.  The particle density for the triangular ansatz is
more peaked at the centre and has a somewhat longer tail at large
distances.

\begin{figure}
\inseps{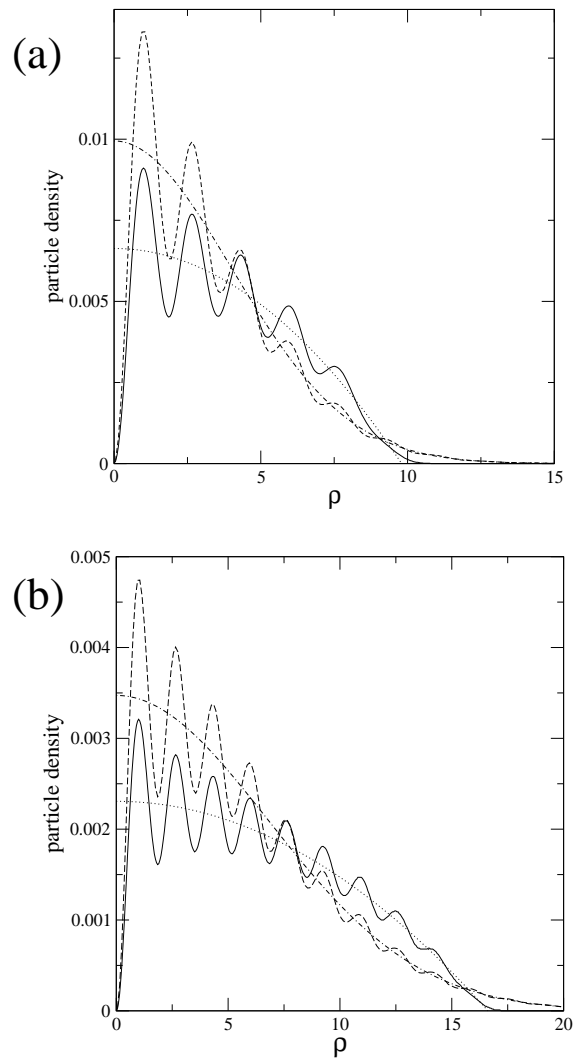}{0.4}
  \caption{\label{fig:3}The angular-averaged particle density as a function of
    the radial co-ordinate for (a) $\ell=31$ and (b) $\ell=91$. (The particle
    density is in units of $a_\perp^{-2}$ and $\rho$ in units of $a_\perp$.)
    In both figures, the solid line corresponds to the full variational
    result, and the dashed line corresponds to the triangular lattice ansatz.
    The dotted lines are the Thomas-Fermi profiles, Eqn. (\ref{eq:tf}), for
    (a) $\ell=31$ and (b) $\ell=91$, while the dot-dashed curves are the
    Gaussian profiles (\ref{eq:gaussian}) for the lattice constant (a)
    $a=1.9351 a_\perp$, and (b) $a=1.9151a_\perp$.}
\end{figure}

To make a more quantitative comparison of the above lattices we have
determined the angular-averaged particle density as a function of the radial
distance $\rho$ from the trap centre, by writing $\eta = \rho e^{i\varphi}$
and averaging the particle density over the azimuthal variable $\varphi$.  The
results are presented in Figs.~\ref{fig:3}(a) and (b).
The difference between the full variational result and the triangular lattice
is very clear for $\ell=31$ as well as for $\ell=91$. For both values of
angular momentum, the average density profile for the full variational result
is very well reproduced by the Thomas-Fermi profile, Eqn. (\ref{eq:tf})
(ignoring the rapid oscillations on the lengthscale of the intervortex
spacing).  On the other hand, the average particle density for the triangular
lattice ansatz is well reproduced by the Gaussian formula (\ref{eq:gaussian}),
consistent with the result of Ref.~\cite{ho}. The Gaussian and
Thomas-Fermi profiles are sufficiently different that it can be clearly seen
that the Thomas-Fermi profile provides a significantly better fit to the full
variational groundstate than does the Gaussian.

\begin{figure} 
  \epsfig{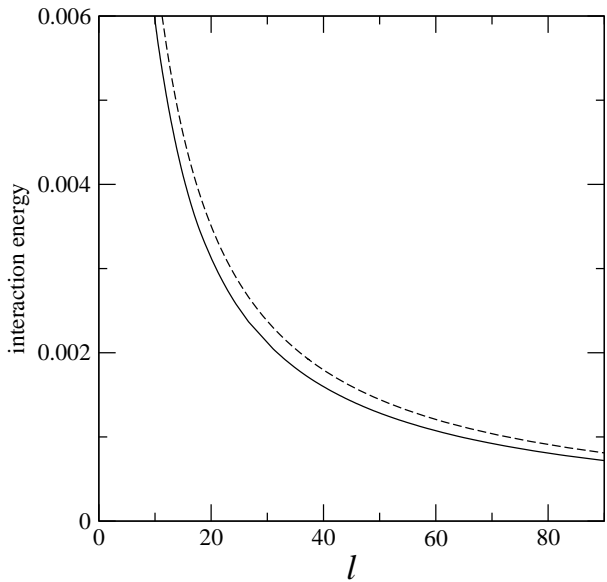}
  \caption{\label{fig:5}The interaction energy (\ref{eq:psi4}) as a
function of the angular momentum per particle, for the full
variational groundstate (solid line) and for the triangular lattice
ansatz (dashed line).  [The interaction energy is plotted in units of
$2\pi\hbar^2 a_s N^2/(Ma_\perp^2 a_\parallel)$.]}
\end{figure}

Finally, in Fig.~\ref{fig:5} we compare the interaction energies per
particle for the full variational wavefunction and the triangular
lattice cases as a function of angular momentum per particle.  As is
required for consistency, the full variational wavefunction has an
interaction energy lower than that of the triangular lattice ansatz
(by about 10\%). The interaction energy for the full variational
wavefunction is very closely given by the expression $E^{\rm var}_{\rm
int} = 2 b g_{2D} N/[9\pi a_\perp^2(\ell+1)]$, which can be found by
use of the Thomas-Fermi analysis of Ref.~\cite{watanabe}; the
corresponding energy for the triangular lattice is $E^{\rm tri}_{\rm
int} = b g_{2D} N/[4\pi a_\perp^2(\ell+1)]$. This energy reduction by
a factor of $8/9$ at fixed angular momentum, $\ell$, is equivalent to
the energy reduction by a factor of $2\sqrt{2}/3$ at fixed angular
frequency $\Omega$ found in Ref.~\cite{watanabe}.

\section{Conclusions}

In conclusion, we have presented the results of numerical calculations
of the finite-size vortex lattices in atomic Bose condensates at large
values of the angular momentum. We have studied the weakly-interacting
limit where vortex cores overlap strongly, and the particles are
restricted to states in the lowest Landau level. Our results show
clearly that the density distribution of particles averaged over a
lengthscale larger than the vortex lattice is very accurately given by
a Thomas-Fermi profile.  These results indicate that, in the lowest
Landau level limit, a fit of the overall density profile found in
experiment to a Thomas-Fermi profile, Eqn.(\ref{eq:tf}), can be used
as a measure of the angular momentum per particle.  Images of the
groundstate density profile and vortex locations can be obtained
reliably from the expanded cloud (following release of the trapping
potential), since for states in the lowest Landau level the
wavefunction of the final (expanded) state is directly related to the
initial (unexpanded) up to a rescaling and rotation \cite{readcooper}.

While preparing this work for publication, we became aware of related
numerical studies where similar conclusions were reached \cite{aps}.

{\bf Acknowledgements:} We are grateful to C. Pethick, L. Radzihovsky
and J. Sinova for helpful correspondence. This work was supported by
EPSRC grant nos. GR/R99027/01 (NRC), GR/R96026/01 (SK) and NSF grant
no. DMR-02-42949 (NR).

% References:

\end{document}